\documentclass[aps,prl,twocolumn,superscriptaddress,amsmath,amssymb]{revtex4-1}

\usepackage{physics}
\usepackage{float}
\usepackage{makeidx}
\usepackage{graphicx,amsmath}
\usepackage{colordvi}
\usepackage[dvipsnames]{xcolor}
\usepackage{ulem}
\usepackage{siunitx}
\usepackage{hyperref}
\usepackage[utf8]{inputenc}
\usepackage[T1]{fontenc}

\usepackage{mathrsfs}

\begin{document}

\title{Ultracold Feshbach molecules in an orbital optical lattice}

\author{Yann Kiefer}
\affiliation{Institut f\"ur Laser-Physik, Universit\"at Hamburg, Luruper Chaussee 149, 22761 Hamburg, Germany}
\affiliation{Zentrum f{\"u}r optische Quantentechnologien, Universit{\"a}t Hamburg, 22761 Hamburg, Germany}

\author{Max Hachmann}
\affiliation{Institut f\"ur Laser-Physik, Universit\"at Hamburg, Luruper Chaussee 149, 22761 Hamburg, Germany}
\affiliation{Zentrum f{\"u}r optische Quantentechnologien, Universit{\"a}t Hamburg, 22761 Hamburg, Germany}

\author{Andreas Hemmerich}
\affiliation{Institut f\"ur Laser-Physik, Universit\"at Hamburg, Luruper Chaussee 149, 22761 Hamburg, Germany}
\affiliation{Zentrum f{\"u}r optische Quantentechnologien, Universit{\"a}t Hamburg, 22761 Hamburg, Germany}
\email{hemmerich@physnet.uni-hamburg.de}
\maketitle

{\bf Quantum gas systems provide a unique experimental platform to study a fundamental paradigm of quantum many-body physics: the crossover between Bose-Einstein condensed (BEC) molecular pairs  and Bardeen Cooper Schrieffer (BCS) superfluidity. Some studies have considered quantum gas samples confined in optical lattices, however, focusing on the case, when only the lowest Bloch band is populated, such that orbital degrees of freedom are excluded. In this work, for the first time, ultracold Feshbach molecules of fermionic $^{40}K$ atoms are selectively prepared in the second Bloch band of an optical square lattice, covering a wide range of interaction strengths including the regime of unitarity. Binding energies and band relaxation dynamics are measured by means of a method resembling mass spectrometry. The longest lifetimes arise for strongly interacting Feshbach molecules at the onset of unitarity with values around $300\,$ms for the lowest band and $100\,$ms for the second band. In the case of strong confinement in a deep lattice potential, we observe bound dimers also for negative values of the $s$-wave scattering length, extending previous findings for molecules in the lowest band. Our work prepares the stage for orbital BEC-BCS crossover physics.}

The crossover between the regimes of BEC and BCS superfluidity is a hallmark of quantum gas physics \cite{Reg1:04, Reg2:04, Zwi:04, Bou:04, Bar:04, Ket:08, Zwe:12, Toe:15}. In most studies, the quantum gas sample is held in a nearly harmonic optical trapping potential. BEC-BCS crossover physics in optical lattices \cite{Lew:07, Gro:17} has been much less explored, and this research is limited to the lowest Bloch band, which exclusively provides local $s$-orbitals. For example, in the ground state of a three-dimensional (3D) optical lattice, binding energies of strongly interacting fermionic potassium pairs have been studied \cite{Sto:06}. Signatures of coherence and superfluidity have been reported \cite{Chi:06} for strongly interacting fermionic lithium pairs. In an earlier work with fermionic potassium, prepared in the transverse ground state of an array of effectively one-dimensional wave guides \cite{Mor:05}, Feshbach dimers have been shown to exist even at negative scattering lengths owing to a confinement induced scattering resonance \cite{Ols:98, Ber:03}. Very recently, $p$-wave interacting atomic pairs tightly confined in excited motional states of isolated microscopic traps have been investigated \cite{Ven:22}. Similarly, as orbital degrees of freedom in electronic condensed matter, e.g. in transition metal oxides \cite{Tok:00, Mae:04}, can give rise to unconventional order, the combination of the conventional BEC-BCS scenario with orbital physics in higher Bloch bands holds the intriguing perspective to discover unexplored fundamental many-body phases such as exotic forms of superfluidity \cite{Li:16}. Examples of chiral, nematic, or topological superfluids have been experimentally demonstrated for bosonic atoms in the second Bloch bands of square \cite{Wir:11, Koc:16}, triangular \cite{Wang:22}, or hexagonal \cite{Wang:21} optical lattices, respectively. Extending these scenarios to composite bosons composed of pairs of fermionic atoms would open up a new regime of unconventional BEC-BCS physics.

\begin{figure}[h]
\centerline{\includegraphics[width=8.6cm]{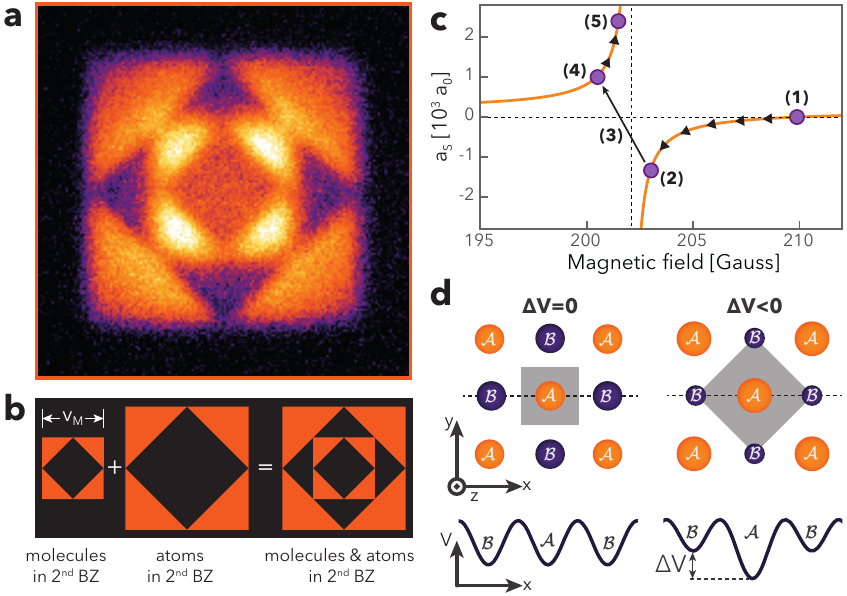}}
\caption{\textit{Production and detection of Feshbach molecules.} (a) A single mass spectrometry image of a mixture of $^{40}$K atoms and $^{40}$K Feshbach dimers excited to the second Bloch band. (b) Composition of the image in (a) by the small second BZ of the molecules and the double-sized second BZ of the atoms. The velocity scale is $v_M =\sqrt{2} \,2\hbar k / M$ with $k=2\pi/\lambda$ and the mass $M$ of the Feshbach dimers. (c) Feshbach resonance and protocol for molecule preparation. See text. (d) Sketch of the lattice geometry in the xy-plane for $\Delta V = 0$ (left) and $\Delta V < 0$ (right). The grey shaded squares denote the corresponding Wigner-Seitz unit cells. At the lower edge, sections along the dashed lines in the upper panels of (d) are shown.}
\label{fig:fig1}
\end{figure}

In a recent work \cite{Hac:21}, we have demonstrated spin-polarized non-interacting fermionic potassium atoms ($^{40}$K) and weakly interacting spin-mixtures in higher Bloch bands of an optical square lattice. The present work investigates the vastly different regime of strong interactions between spin-up and spin-down fermions accessed by tuning a Feshbach resonance. This allows us to demonstrate and investigate bosonic Feshbach dimers \cite{Chi:10}, composed of fermionic $^{40}$K atoms, in higher Bloch bands of an optical square lattice. We study shallow and deep lattices, where the dimers can tunnel or are confined in one-dimensional channels. A method inspired by mass spectrometry is applied to discriminate atoms and dimers by separating them in a ballistic time-of-flight protocol (see Methods). This leads to images of the Brillouin zone (BZ) structures of atoms and molecules in velocity space, which differ in size by their mass ratio. This is seen in Fig.~\ref{fig:fig1}(a) for a mixture of molecules and atoms populating the second Bloch band. The molecules appear in the small - and the atoms in the large second BZ, forming a nested structure according to the sketch in Fig.~\ref{fig:fig1}(b). Hence, atoms and molecules can be separately counted. Using this method, we present measurements of binding energies, dissociation dynamics, and exceptionally long molecule lifetimes for a wide range of interaction strengths including the strongly correlated regime. Such long lifetimes exceeding all other relevant time scales are crucial, to study long-lived equilibrium states of the system.

\textit{Production of ultracold Feshbach molecules.}
In short, the production of Feshbach dimers in the second Bloch band of a square lattice proceeds as follows: A balanced fermionic spin mixture is produced in the lowest Bloch band of the lattice. Subsequently, spin-up and spin-down atoms are associated to form Feshbach dimers by rapid adiabatic tuning of the magnetic field across an $s$-wave Feshbach resonance. Finally, a quench of the lattice potential selectively excites the dimers to the second Bloch band. 

The following detailed protocol is applied. A non-interacting spin-polarized degenerate Fermi gas of $6 \times 10^5$ $^{40}$K atoms in the state $\ket{F=9/2,m_F=9/2}$ at a temperature of $T = 0.17 \,T_F$ is prepared in an optical dipole trap (ODT), formed by two orthogonally intersecting laser beams with a wavelength of $\lambda=1064\,$nm, where $T_F$ denotes the Fermi temperature. Note that for spin mixtures prepared at low magnetic fields, the lowest temperature reached in the ODT is $T/T_F=0.09$. Next, a radio-frequency with a constant value of $46\,$MHz is applied, while a homogeneous magnetic field $B$ (pointing along the $z$-axis) is ramped up from zero to approximately $B_0 \approx 209.9\,$G, such that a rapid adiabatic passage is obtained that inverts the sample to the $\ket{\downarrow} \equiv$ $\ket{F=9/2,m_F=-9/2}$ state. Due to a Feshbach resonance located at $B_{\textrm{res}} = 202.1\,$G \cite{Reg1:04}, the $s$-wave scattering length for contact interaction between the states $\ket{\downarrow}$ and $\ket{\uparrow} \equiv$ $\ket{F=9/2,m_F=-7/2}$, approximated as
\begin{equation}
\begin{aligned} 
a_S(B)=a_{bg}\,[1-\Delta B/(B-B_{\textrm{res}})]\, ,
\end{aligned}
\label{eq:FBR}
\end{equation}
takes the value $a_S(B_0) \approx 0$, which corresponds to position (1) in Fig.~\ref{fig:fig1} (c). Here, $a_{bg}=174\,a_0$ is the background scattering length and $\Delta B = 7.8\,$G is the width of the Feshbach resonance. Subsequently, the sample is adiabatically loaded into the lowest Bloch band of a bipartite optical square lattice, formed by two mutually orthogonal standing waves with wavelengths $\lambda =1064\,$nm, oriented perpendicularly to the $z$-axis. The optical lattice is shaped in a Michelson-Sagnac interferometer that allows for precise control of the associated band structure (for details see Ref.~\cite{Hac:21}). The resulting optical potential is composed of two classes of independently tunable potential wells $\mathcal{A}$ and $\mathcal{B}$ arranged as the black and white squares of a chequerboard (see Fig.~\ref{fig:fig1} (d)). In the $xy$-plane, the lattice potential can be approximated by
\begin{equation}
\begin{aligned} 
V(x, y)=&-V_{0}\left[\cos ^{2}(k x)+\cos ^{2}(k y)\right] \\ 
&-\frac{1}{2} \Delta V \cos (k x) \cos (k y) 
\end{aligned}
\end{equation}
with the wave number $k=2\pi/\lambda$, the lattice depth parameter $V_0$, and $\Delta V \equiv - 4 V_0 \cos(\theta)$ denoting the potential difference between the $\mathcal{A}$ and $\mathcal{B}$ wells. The experimental parameter $\theta$ can be tuned within the interval $[0,\pi]$, i.e., $\Delta V \in 4 V_0 \times [-1,1]$. Note that tightly bound dimers with mass $M \equiv 2m$ possess twice the polarizability and hence twice the value of $V_0$ as compared to atoms with mass $m$. Henceforth, we indicate lattice depth parameter values for atoms and molecules as $V_0^{(m)}$ and $V_0^{(M)} =2\,V_0^{(m)}$, respectively. Along the $z$ direction, the atoms are held by the weak approximately harmonic confinement of the optical dipole trap, such that the lattice wells acquire a tubular shape. After lattice loading, we apply a radio frequency pulse during $13\,\mu$s to create a Fermi gas with equal populations in the states $\ket{\uparrow}$ and $\ket{\downarrow}$ at position (1) in Fig.~\ref{fig:fig1}(c). Next, by tuning $B$ to $202.3\,$G (cf. position (2) in Fig.~\ref{fig:fig1} (c)), the $s$-wave scattering length $a_S$ for collisions between $\ket{\uparrow}$ and $\ket{\downarrow}$ is adjusted to a large negative value, in order to obtain efficient evaporative cooling of the atomic sample in presence of the lattice. Finally, the previously unpaired mixture of Fermions is converted into bosonic molecules by adiabatically sweeping the magnetic field across the Feshbach resonance from $202.3\,$G to $200.46\,$G (indicated as step (3) in Fig.~\ref{fig:fig1}(c)). For the lowest temperatures, upon arrival at position (4) in Fig.~\ref{fig:fig1}(c), we observe close to hundred percent conversion efficiencies with no discernible atomic fraction. Next, a quench from an initial value $\theta \approx 0.4\, \pi$, used for loading the lowest band, to $\theta \approx 0.53\,\pi$ efficiently prepares a large fraction of the atoms or molecules in the second band (for details see Ref. \cite{Hac:21}). By means of a final adiabatic change of $B$ (position (5) in Fig.~\ref{fig:fig1} (c)), we may subsequently tune to a desired target value of $a_S$, which lets us adjust the molecular binding energy.  

\begin{figure}
\centerline{\includegraphics[width=9cm]{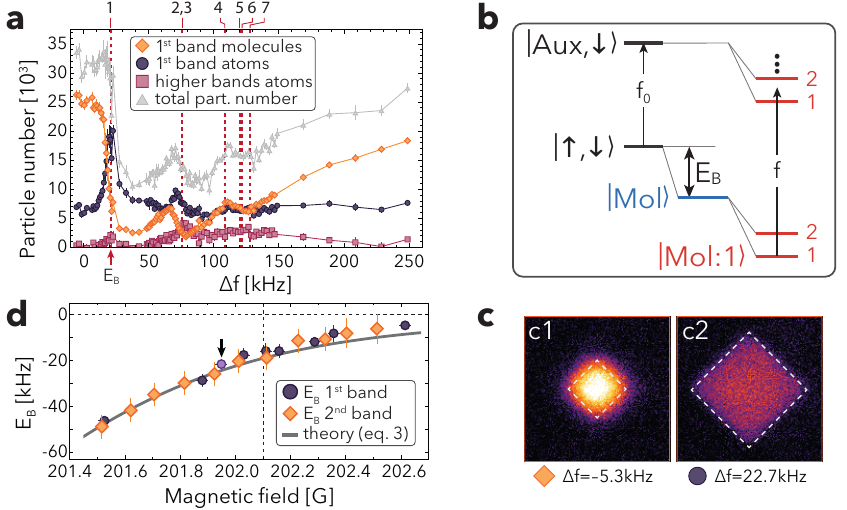}}
\caption{\textit{Binding energies in a deep lattice.} (a) A typical dissociation spectrum for a magnetic field $B = 201.95\,$G, i.e., corresponding to the data point in (d) highlighted by an arrow, with molecules initially prepared in the lowest Bloch band ($V_0^{(M)} = 40\,E_\textrm{rec}^{(M)}$, $\theta = 0.4 \,\pi$). The orange diamonds and dark purple disks show the molecule and atom populations in the first band, respectively, plotted against $\Delta f = f - f_0$, with the irradiated radio frequency $f$, and $f_0$ according to (b). The leftmost vertical red dashed line indicates the frequency, where the first significant drop of the molecule population and a corresponding peak of the atom population in the first band is observed. This frequency is identified with the molecular binding energy $E_B$. Further dashed vertical lines denote the positions of higher Bloch bands $\ket{\textrm{Aux},\downarrow \, : \nu}$ with $\nu \in \{2,3,4,5,6,7\}$, leading to further resonant molecule dissociation (cf. (b)). The error bars show the standard deviation of the mean for $15$ experimental runs. (c) Mass spectrometry images for $\Delta f = -5.3\,$kHz (c1) and $\Delta f = 22.7\,$kHz (c2). The white dashed rectangles show the first BZ for molecules (c1) and atoms (c2). (d) Measured binding energies for molecules prepared in the first (dark purple disks) and second (orange diamonds) Bloch band with ($V_0^{(M)} = 40\,E_\textrm{rec}^{(M)}$, $\theta = 0.4 \,\pi$) and ($V_0^{(M)} = 60\,E_\textrm{rec}^{(M)}$, $\theta = 0.54 \,\pi$), respectively. The error bars are estimated from the spectral width of the underlying dissociation resonance. The dashed vertical line indicates the position of Feshbach resonance $B_{\textrm{res}}$. The grey line shows a calculation using Eq.~\ref{eq:CIR}.}
\label{fig:fig2}
\end{figure}

\textit{Binding energies} 
Binding energy of Feshbach molecules are measured by dissociating them with a $5\,$ms long radio frequency pulse, converting $\ket{\uparrow}$ atoms into atoms in the auxiliary state $\ket{\textrm{Aux}} \equiv \ket{F=9/2, m_F=-5/2}$. The unbound $\ket{\downarrow}$ atoms, which remain trapped in the optical lattice, are readily discriminated from the molecules as explained in the context of Fig.~\ref{fig:fig1} (a,b) and in Methods. 

Let us first discuss the case of a deep optical lattice, such that tunnelling is practically suppressed for molecules and their motion is restricted to quasi one-dimensional (1D) tubes along the $z$-direction. An example for this regime is realized by adjusting $V_{0}^{(M)} = 40\,E_{\textrm{rec}}^{(M)}$, $\theta=0.4\,\pi$ and hence $\Delta V^{(M)} = -49.4\,E_{\textrm{rec}}^{(M)}$ with $E_{\textrm{rec}}^{(M)} \equiv \frac{\hbar^2k^2}{2 M}$ denoting the single photon recoil energy for molecules. A typical example of a dissociation spectrum for molecules $\ket{\textrm{Mol}:1}$ prepared in the lowest band is shown in Fig.~\ref{fig:fig2}(a) for $B = 201.95\,$G corresponding to the data point in (d) indicated by an arrow. We plot the numbers of molecules (diamonds) and atoms (disks) in the lowest band, the atoms in all excited bands (squares), and the total number of particles, i.e., atoms plus molecules (triangles), against $\Delta f = f - f_0$, i.e., the applied radio frequency $f$ minus the $\ket{\uparrow} \rightarrow \ket{\textrm{Aux}}$ transition frequency $f_0$ (see Fig.~\ref{fig:fig2}(b)). The latter depends only on $B$ and is readily measured after preparing a pure $\ket{\uparrow}$ sample in the dipole trap. Hence, $\Delta f = 0$ corresponds to zero molecular binding energy.

To understand the spectral features in Fig.~\ref{fig:fig2}(a), it is helpful to first look at the atomic and molecular levels sketched in Fig.~\ref{fig:fig2}(b). The black horizontal bars show the energies of the bare two-atom states $\ket{\uparrow,\downarrow}$ and $\ket{\textrm{Aux},\downarrow}$ separated by the frequency $f_0$. The blue horizontal bar shows the energy of the Feshbach molecules $\ket{\textrm{Mol}}$ made of atom pairs $\ket{\uparrow,\downarrow}$, shifted by the binding energy $E_B$. The red horizontal bars show the additional energy shifts for the Bloch bands due to the presence of the optical lattice. The molecules $\ket{\textrm{Mol}}$ are prepared in the lowest molecular band, denoted $\ket{\textrm{Mol}:1}$ in Fig.~\ref{fig:fig2}(b). As the radio frequency $f$ is increased, we expect the first drop of the molecular population in $\ket{\textrm{Mol}:1}$, when $f$ reaches the resonance frequency $f_1$ for the transition $\ket{\textrm{Mol}:1} \rightarrow \ket{\textrm{Aux},\downarrow \,: 1}$ such that unbound atoms in the first Bloch band are produced. According to Fig.~\ref{fig:fig2}(a), this occurs at $\Delta f = \Delta f_1 \equiv 22.7\,$kHz, which is identified with the value of the binding energy $E_B$. To further illustrate the conversion of molecules into atoms, mass spectrometry images are shown in Fig.~\ref{fig:fig2}(c) for $\Delta f = -5.3\,$kHz, well below $\Delta f_1$ (c1), and at $\Delta f = \Delta f_1$ (c2). In fact, in (c1), predominantly molecules in the first BZ are seen, while in (c2) most molecules are dissociated into atoms, giving rise to a first BZ expanded by a factor two.

At larger values of $\Delta f  > \Delta f_1$, further resonances occur, leading to reduced molecule numbers due to dissociation into higher bands $\ket{\textrm{Aux},\downarrow \, : \nu}$ with $\nu \in \{2,3,5,6,7\}$. The respective resonance frequencies are readily calculated by a band structure calculation and are plotted into Fig.~\ref{fig:fig2}(a) as vertical dashed red lines. Note that, for the fourth band $\nu = 4$, no resonance arises, which can be explained by the small Franck-Condon overlap between the wave function of $\ket{\textrm{Mol}:1}$ and $\ket{\textrm{Aux},\downarrow \, :4 }$. This has been checked by calculating the respective Bloch functions, with the result that $\ket{\textrm{Mol}:1}$ predominantly resides in the deep wells and $\ket{\textrm{Aux},\downarrow \, :4}$ in the shallow wells.

Dissociation spectra as in Fig.~\ref{fig:fig2}(a) let us determine the binding energies for molecules prepared in the first (dark purple disks) and second (orange diamonds) bands, shown in Fig.~\ref{fig:fig2}(d). The gray solid line shows a calculation using the implicit equation
\begin{equation}
\frac{a_S - r_0}{(1+\delta) a_{r}}=-\frac{1}{\zeta\left(1 / 2, -E_{B} / 2 \hbar \omega_{r}\right)}.
\label{eq:CIR}
\end{equation}
adapted from Ref.~\cite{Ber:03} for a single 1D tubular potential, where $\zeta$ denotes the Hurwitz zeta function. We insert the radial fundamental frequency $\omega_r$, measured in the radially symmetric quasi 1D tubes of our lattice, and the associated harmonic oscillator length $a_{r}\equiv \sqrt{\hbar/ \mu \omega_r}$ for the relative atomic motion with $\mu$ denoting the reduced atomic mass. We multiply $a_{r}$ by $1+\delta$ with a small positive $0<\delta \ll 1$, to account for the anharmonicity of the tube potential. The free space $s$-wave scattering length $a_S$ is expressed as a function of the magnetic field $B$ according to Eq.~\ref{eq:FBR}. Finally, following Ref.~\cite{Gri:93}, to account for a realistic van der Waals scattering potential $-C_6\,r^{-6}$, we introduce the finite range parameter $r_0 = (m C_6 / 32 \hbar^2)^{1/4} \, \Gamma(3/4)/ \Gamma(5/4) $, and replace $a_S$ by $a_S - r_0$. From Refs.~\cite{Tic:04, Fal:08} we take $C_6 = 3926\,a_0$ and $r_0 = 65\,a_0$. Note that Eq.~\ref{eq:CIR} is configured such that for zero interaction, zero binding energy is obtained. This theoretical model, based on Refs.~\cite{Gri:93, Ols:98, Ber:03}, reproduces the binding energies in Fig.~\ref{fig:fig2}(d) remarkably well if one sets $\delta = 0.142$, which reasonably well agrees with the expected anharmonicity. The model accounts for effects of reduced dimensionality arising if the degree of radial confinement becomes comparable with $a_S$, giving rise to a confinement induced resonance of the effective 1D scattering cross section. As a consequence, bound states become possible for negative values of $a_S$ as is seen in Fig.~\ref{fig:fig2}(d). This has been previously reported for potassium Feshbach molecules in the lowest transverse state of an array of isolated 1D optical traps in Ref.~\cite{Mor:05}. Note that Fig.~\ref{fig:fig2}(d) shows equal binding energies for molecules prepared in the first and second bands. This results from engineering the lattice potentials via adjustment of $V_0$ and $\Delta V$, such that the lattice wells with predominant population, i.e., the deep wells, if the molecules are prepared in the first and the shallow wells if prepared in the second band, respectively, exhibit equal values of $\omega_r$ and $a_{r}$. For example, with $V_{0}^{(M)} = 40\,E_{\textrm{rec}}^{(M)}$ and $\theta = 0.4 \,\pi$ for molecules in the first band, and $V_{0}^{(M)} = 60\,E_{\textrm{rec}}^{(M)}$ and $\theta = 0.54 \,\pi$ for molecules in the second band, as used in Fig.~\ref{fig:fig2}(d), $\omega_r = 2 \pi \, \times 30\,$kHz is obtained.

\begin{figure}
\centerline{\includegraphics[width=8.6cm]{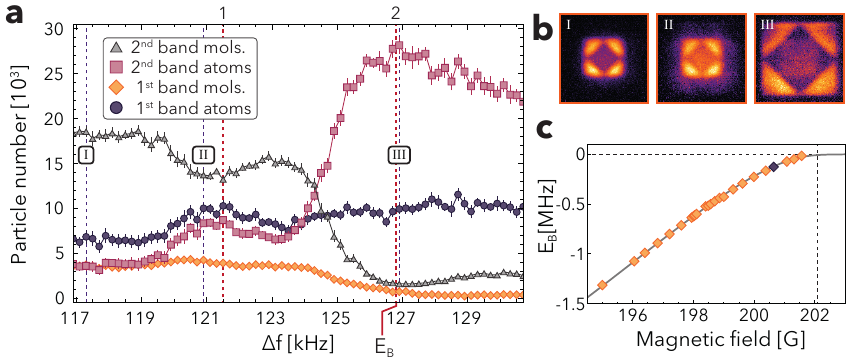}}
\caption{\textit{Binding energies in a shallow lattice.} (a) Dissociation spectrum for molecules initially prepared in the second band with a lattice depth parameter $V_{0}^{(M)} = 10\,E_{\textrm{rec}}^{(M)}$ and $\theta = 0.54\,\pi$. The magnetic field is $B=200.6\,$G corresponding to the dark purple diamond in (c). Orange diamonds (dark purple disks) show populations of molecules (atoms) in the first band. Grey triangles (magenta squares) show populations of molecules (atoms) in the second band. The red dashed lines indicate the two frequencies $\Delta f_1$ and $\Delta f_2$, where dissociation maximally depletes the population of molecules in the second band. The error bars show the standard deviation of the mean for $15$ experimental runs. (b) Mass spectrometry images for the values of $\Delta f$ indicated by I, II, III in (a). The orange diamonds in (c) show binding energies obtained from spectra as plotted in (a) for varying magnetic fields. The errors are estimated from the spectral width of the underlying dissociation resonance to be less than $5\,$kHz, i.e. an order of magnitude smaller than the data symbols. The grey solid line presents the theoretical prediction of Eq.~\ref{eq:CIR}.} 
\label{fig:fig3}
\end{figure}

Next, we discuss molecules prepared in the second band ($\ket{\textrm{Mol} : 2}$) for a shallow lattice with $V_{0}^{(M)} = 10\,E_{\textrm{rec}}^{(M)}$, $\theta = 0.54\,\pi$ and hence $\Delta V^{(M)} = 5 \,E_{\textrm{rec}}^{(M)}$ for an extended range of the magnetic field below the Feshbach resonance, $B < B_{\textrm{res}}$. In Fig.~\ref{fig:fig3}(a), we show a dissociation spectrum for molecules at $B=200.6\,$G corresponding to the dark purple diamond-shaped data point in (c). The plot in (a) shows the second band molecules (grey triangles), whose number is initially maximized, second band atoms (magenta squares), and molecules (orange diamonds) and atoms (dark purple disks) in the first band. The red dashed lines emphasize the frequencies $\Delta f_1$ and $\Delta f_2$, where local minima in the number of molecules in the second band are found, due to maximal efficiency of the dissociation process. At frequencies $\Delta f \ll \Delta f_1$ dissociation is not resonant such that in mass spectrometry images one observes the second BZ filled with molecules, as exemplified in the panel (I) of Fig.~\ref{fig:fig3}(b), recorded at position (I) in (a). At the left red dashed line ($\Delta f_1$), dissociation arises due to a transition, mainly exciting second band molecules $\ket{\textrm{Mol} : 2}$ to atomic pairs $\ket{\textrm{Aux},\downarrow \,: 1}$ in the first band. This is confirmed by the mass spectrometry image in panel (II) in (b), recorded at position (II) in (a), close to the left red dashed line. Here, a partial filling of the first BZ with atoms is observed. The dissociation process shows limited efficiency due to the small Franck Condon overlap between the wave functions involved, similarly as discussed in the context of Fig.~\ref{fig:fig2}(a) for the transition $\ket{\textrm{Mol} : 1} \rightarrow \ket{\textrm{Aux},\downarrow \,: 4}$. Around the second red dashed line ($\Delta f_2$), the dissociation transition couples $\ket{\textrm{Mol} : 2}$ to $\ket{\textrm{Aux},\downarrow \,: 2}$. Both states belong to second bands with a sizable Franck Condon overlap, such that the dissociation around $\Delta f_2$ is notably more efficient than for $\Delta f_1$, as seen by the nearly complete depletion of the molecule population in (a) around $127\,$kHz. Panel (III) in Fig.~\ref{fig:fig3}(b) confirms that the dissociated atoms in fact arise in the second band, giving rise to a filled second BZ for atoms, twofold increased as compared to the second BZ for molecules in panel (I). Note that $\Delta f_2 -\Delta f_1$ is approximately given by the separation of the second and first bands for atoms. A determination of $\Delta f_1, \Delta f_2$ below the scale of a few kHz is not supported by the width of the spectral features observed near the red dashed lines in Fig.~\ref{fig:fig3}(a). In Fig.~\ref{fig:fig3}(c), binding energies for molecules in the second band are shown, measured by analyzing dissociation spectra as in (a) for different magnetic fields. The grey solid line represents the theoretical prediction according to Eq.~\ref{eq:CIR}, showing remarkable agreement.

\textit{Molecular decay dynamics.} In this section we discuss the observation of two relaxation channels for Feshbach dimers \cite{Shi:11}. The first process dominates for $\xi \equiv (k_F a_S)^{-1} \gg 1$ ($k_F \equiv$ Fermi momentum), i.e., for relatively weak scattering lengths, where Feshbach dimers become deeply bound. In this regime, the primary relaxation process is based on inelastic dimer-dimer collisions, where one molecule gains binding energy, while the other is dissociated, such that both molecules are lost. Dimer-atom collisions are less relevant since our experiments start with a nearly pure molecule sample. Larger binding energies provide larger Franck-Condon overlap between the involved molecular wave functions, so that the molecular lifetime $\tau$ should decrease with $\xi$, in accordance with the scaling $\tau \propto (a_S)^{2.55}$ predicted in absence of a lattice \cite{Pet:05}. In the other extreme for $\xi \ll 1$, according to Ref.~\cite{Pet:05}, the elastic dimer-dimer scattering length $a_{dd} $ scales as $a_{dd} \approx 0.6\,a_S$ and hence due to the large size of $a_S$, molecular three-body collisions are expected to introduce molecule loss and to give rise to a lifetime that decreases, when $\xi$ approaches zero.

\begin{figure}
\centerline{\includegraphics[width=9.0cm]{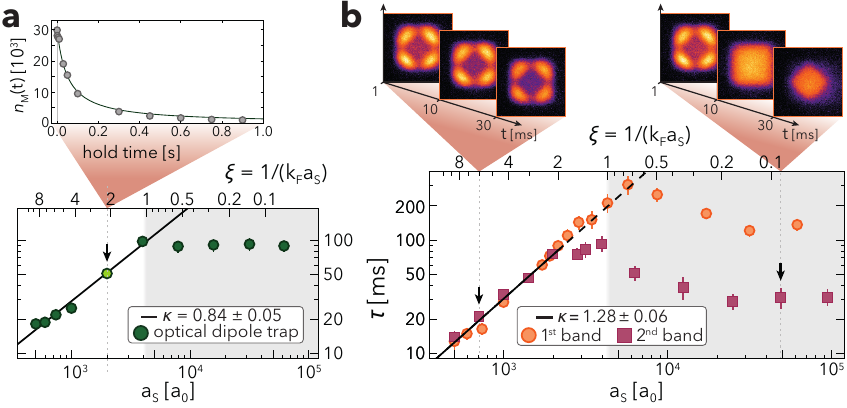}}
\caption{\textit{Relaxation dynamics.} (a) Half-lives $\tau$ of Feshbach dimers in the dipole trap plotted versus $\xi$ (upper axis) and $a_S$ (lower axis). The inset shows an exemplary measurement of $\tau$ for the light green data point indicated by the black arrow. The black solid line is a fit for the domain $\xi >2$ with a straight line, extrapolated to the domain $\xi < 2$, indicating a power law $\propto \xi^{-\kappa}$ with $\kappa = 0.84 \pm 0.05$. The orange disks and magenta squares in (b) show half-lives of Feshbach dimers prepared in the first and second Bloch bands of an optical lattice, respectively. The lattice depth is $V_{0}^{(M)} = 8\,E_{\textrm{rec}}^{(M)}$ and $\theta = 0.4\,\pi$ for the first Bloch band and $0.535\,\pi$ for the second Bloch band. The black solid line is a common fit for data of both bands within the domain $\xi >2$ with a straight line, extrapolated to the domain $\xi < 2$ (dashed continuation), indicating a power law $\propto \xi^{-\kappa}$ with $\kappa = 1.28 \pm 0.06$. The error bars for $\tau = (n_0 \, \beta)^{-1}$ in (a) and (b) are determined by propagating the errors for $n_0$ and $\beta$ obtained in the fit procedure. The insets at the top boundary of (b) illustrate the different relaxation paths found in the second Bloch band for small and large $\xi$ (cf. text).}
\label{fig:fig4}
\end{figure}

In Fig.~\ref{fig:fig4}(a), we benchmark the lifetimes of the Feshbach molecules trapped in the approximately harmonic optical dipole trap. After the dimers are formed (Fig.~\ref{fig:fig1}(c)), the scattering length $a_S$ is adjusted to a desired value via the magnetic field according to Eq.~\ref{eq:FBR} and the dimers are held for a variable time in the trap. In order to count the remaining number of molecules $n_{M}$, the magnetic field is rapidly tuned to $B_{\textrm{img}}=200.46\,$G, associated with a moderate value of $a_S$ and the molecules are allowed to ballistically expand (nearly unimpaired by interaction) during $22\,$ms, before an absorption image of the $\ket{\uparrow}$ atoms is recorded. The dimer population plotted against the hold time is fitted with the two-body decay model $\dot{n}_{M} = -\beta \,n_{M}^2$, with the solution $n(t) = n_0 \cdot (1 + t / \tau)^{-1}$, the initial number of molecules $n_0 \equiv n(0)$, and the half-time of the molecule sample $\tau \equiv (n_0 \beta)^{-1}$. An exponential fit, assuming density independent loss, or a three-body decay model fail to describe the data. The obtained half-time $\tau$ is plotted in Fig.~\ref{fig:fig4}(a) versus $\xi$ and $a_S$. The inset shows an exemplary measurement of $\beta$ and $n_0$ leading to the light green data point indicated by a black arrow. The main panel shows strikingly different behaviour in the unitarity regime $\xi <1$, where a nearly constant half-time $\tau$ around $100\,$ms is observed, and in the regime $\xi >1$, where the data are well fitted with a straight line, indicating a power law $\tau \propto \xi^{-\kappa}$ with an exponent $\kappa = 0.84 \pm 0.05$. The largest values of $\tau$ are found at $\xi \approx 1$. Note that, $\kappa$ does not agree with the prediction $2.55$ for dimer-dimer collisions in Ref.~\cite{Pet:05} or the experimental value $\approx 2.3$ for mixed dimer-dimer and dimer-atom collisions, reported in Ref.~\cite{Reg2:04}.

In Fig.~\ref{fig:fig4}(b), an analogous analysis is carried out in presence of a shallow optical lattice with $V_{0}^{(M)} = 8\,E_{\textrm{rec}}^{(M)}$, which permits nearest-neighbour tunneling on a sub-millisecond time scale. The orange disks show the observed half-lives for molecules prepared in the first Bloch band with $\theta = 0.4 \,\pi$ and hence $\Delta V_{0}^{(M)} = -9.88\,E_{\textrm{rec}}^{(M)}$. The magenta squares correspond to $\theta = 0.535 \,\pi$ (i.e., $\Delta V_{0}^{(M)} = 3.51\,E_{\textrm{rec}}^{(M)}$) for molecules prepared in the second Bloch band. In the regime $\xi > 2$, both data sets show the same dependence on $\xi$ and are fitted with the same straight line (black solid line in (b)), indicating a power law model $\propto \xi^{-\kappa}$ with $\kappa = 1.28 \pm 0.06$. The dashed black line is a continuation into the $\xi < 2$ domain. Note that also in the presence of a lattice, $\kappa$ does not agree with the prediction $2.55$ in Ref.~\cite{Pet:05} or the experimental value $\approx 2.3$ from Ref.~\cite{Reg2:04}, both obtained for a scenario without a lattice. For $\xi < 1$, we find that $\tau$ grows with increasing $\xi$. While the data for the second band very well agree with the results for the first band in the entire range $\xi > 2$, for $\xi < 2$ a dramatic decrease of $\tau$ is observed for the second band, which only rises up to a threefold shorter lifetime than observed for the first band. This indicates that an additional relaxation channel opens for the second band if $\xi < 2$. In fact, as illustrated in the two insets at the upper edge of Fig.~\ref{fig:fig4}(b), for $\xi$ close to zero, e.g. for $\xi \approx 0.1$, we see an initial pronounced decay of molecules into the first band, followed by a subsequent decay of the first band population, while in the domain $\xi >2$, e.g., close to $\xi \approx 6$, the molecules do not initially transit to the first band during relaxation. This may be explained as follows: For small $\xi$, i.e., large $a_S$, in the second band, the increasing elastic binary collision cross section can lead to a redistribution of energy from the lattice plane into the tube direction, which gives rise to an additional escape channel from the lattice. A similar relaxation channel for weakly interacting bosonic atoms has been recently reported in Ref.~\cite{Nus:20}. Also effects of reduced dimensionality in the tubular lattice sites are expected to be larger in the first than in the second band, which may also contribute to explain the longer lifetimes of first band dimers \cite{Ven:22}.

In summary, our work demonstrates strongly correlated ultracold Feshbach dimers prepared in higher Bloch bands of an optical lattice, which give rise to orbital physics. Using a method reminiscent of mass spectrometry, we find surprisingly long molecular lifetimes in the unitarity regime on the order of hundred milliseconds and study binding energies and dissociation and relaxation dynamics. Our work prepares the stage for future studies of BEC-BCS cross over physics with orbital degrees of freedom.

\begin{figure}[t]
\centerline{\includegraphics[width=5.0cm]{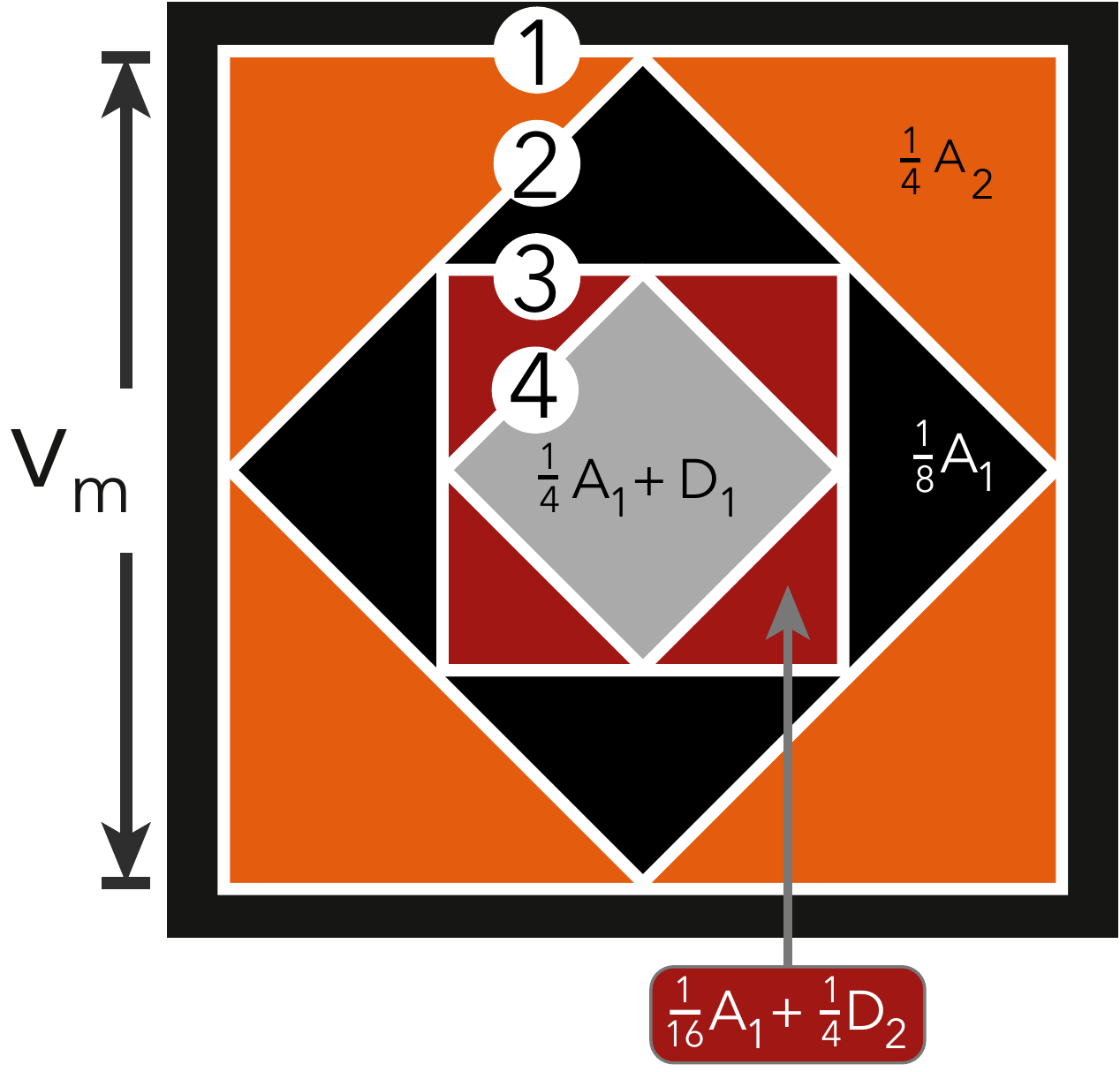}}
\caption{\textit{Evaluation of atom and dimer populations.} $xy$-plane of velocity space with $v_m =\sqrt{2} \,2\hbar k / m$. $A_{\nu}$ and $D_{\mu}$ denote the total atom and dimer populations, respectively, within the Bloch bands with band index $\nu,\mu \in \{1,2\}$. The terms $a_{\nu} A_{\nu} + d_{\mu} D_{\mu}$, plotted inside different sectors bordered by white lines, specify the respective total particle numbers. See text.}
\label{fig:figM}
\end{figure}

\vspace{8 pt}
\noindent \textbf{Methods}
\newline \newline
\noindent \textit{Mass spectrometry protocol}
In order to uniquely distinguish and selectively count the populations of atoms and Feshbach dimers in presence of an optical lattice, they are separated in a ballistic time-of-flight protocol, which maps the population of the $n$-th band to the $n$-th Brillouin zone (BZ). This protocol consists of a rapid adiabatic termination of the lattice potential (in $5\,$ms) followed by a ballistic expansion (in $22\,$ms). During the second half of the expansion the magnetic field is tuned to $B_{\textrm{img}}= 200.6\,$G in order to enable absorption imaging at the same resonant frequency in each experimental run. For the case that the dimer mass $M$ is twice that of an atom $m$, atoms travel twice as fast compared to homonuclear dimers with the same initial momentum. Hence, this protocol arranges the atoms and molecules in velocity space with a BZ structure scaled up by a factor two for atoms as compared to dimers. This is sketched in Fig.~\ref{fig:figM}. The figure shows the first and second BZs for atoms in the background (black and orange areas) and for molecules in the foreground (grey and red areas), giving rise to a nested structure of four white squares denoted 1-4. The optical densities integrated across the areas enclosed by these squares are denoted as $\Box_{\nu}, \nu \in \{1-4\}$. We denote the total populations as $A_{\nu}$ and $D_{\nu}$ for atoms and molecules in the band with band index $\nu \in \{1,2\}$, respectively. If sectors in Fig.~\ref{fig:figM} comprise atoms and dimers, the counting protocol requires the assumption that the first band for atoms is uniformly filled. This assumption is reasonable since, in the experiments described in the main text, we typically start with the lowest band filled with fermionic atoms, form Feshbach dimers, and subsequently excite a fraction of them to the second band without changing their quasi-momenta \cite{Hac:21}. Note that due to the bosonic nature of the Feshbach dimers, the respective Brillouin zones (red and grey areas in Fig.~\ref{fig:figM}) are not necessarily filled homogeneously, since Bose-enhancement and hence multiple occupancy of available energy states are possible. Under the condition of a homogeneous filling of the first atomic band, the populations $A_1, A_2, D_1, D_2$ in different sectors of Fig.~\ref{fig:figM} are indicated in the figure accounting for the four-fold rotation symmetry of the BZs and the four times larger BZ areas for atoms. This leads to the following relations: $A_2 = \Box_1 - \Box_2$, $A_1= 2 (\Box_2 - \Box_3)$, $D_2 = \frac{3}{2} \Box_3 - \Box_4 - \frac{1}{2} \Box_2$, and $D_1 = \frac{1}{2} \Box_3 - \frac{1}{2} \Box_2 + \Box_4$. If all atoms and molecules reside in the second Band, i.e. $A_1= D_1 = 0$, the simple case $A_2 = \Box_1 - \Box_2$ and $D_2 = \Box_3 - \Box_4$ arises, in which case the condition of uniform filling is not required.
\newline \newline 
\textrm{} \vspace{1 pt}
 
\noindent \textbf{Acknowledgments}
We thank Raphael Eichberger for help in the early stage of the experiment. We acknowledge support from the Deutsche Forschungsgemeinschaft (DFG) through the collaborative research center SFB 925 (Project No. 170620586, C1). M.H. was partially supported by the Cluster of Excellence CUI: Advanced Imaging of Matter of the Deutsche Forschungsgemeinschaft (DFG) - EXC 2056 - project ID 390715994.
\\  \\

\end{document}